\documentclass{jpsj2}
\title
{A Mechanism of Spin-Triplet Superconductivity in Hubbard Model on Triangular Lattice: Application to ${\rm UNi_{2}Al_{3}}$}

\author
{ 
Yunori  {\sc Nisikawa}\footnote{E-mail
address:nisikawa@scphys.kyoto-u.ac.jp} and Kosaku {\sc Yamada}
}

\inst
{
Department of Physics, Kyoto University, Kyoto 606-8502}

\abst{We discuss the possibility of spin-triplet superconductivity 
in a two-dimensional Hubbard model on a triangular 
lattice within the third-order perturbation theory. 
When we vary the symmetry in the dispersion of the bare energy band 
from D$_2$ to D$_6$, 
spin-singlet superconductivity in the 
D$_2$-symmetric system  is suppressed and 
we obtain spin-triplet superconductivity in near the D$_6$-symmetric system.
In this case, it is found that the vertex terms, 
which are not included in the interaction mediated by the spin
fluctuation, are essential for realizing the spin-triplet pairing.
We point out the possibility that obtained results correspond to the
difference between the superconductivity of UNi$_2$Al$_3$ and that of 
UPd$_2$Al$_3$.}

\kword{spin-triplet superconductivity, two-dimensional Hubbard model on
triangular lattice, third-order perturbation theory, vertex correction,
UNi$_2$Al$_3$, UPd$_2$Al$_3$}

\begin{document}
\maketitle

Heavy-fermion superconductors have attracted much interest since
 spin-triplet superconductivity has been reported in these compounds.
Respective isostructual hexagonal PrNi$_2$Al$_3$-type
heavy-fermion compounds, UPd$_2$Al$_3$ and UNi$_2$Al$_3$, were found to
be superconductors with $T_{{\rm c}}\simeq 2$ and $1$K, respectively.
Their superconducting states coexist with antiferromagnetic long-range 
orders with $T_{{\rm N}}=14.3$ and $4.5$K, respectively.
The properties of magnetism and superconductivity 
are different in each compound.
UPd$_2$Al$_3$ orders in an antiferromagnetic 
long-range structure with a commensurate wave vector
${\bf k}=(0,0,0.5)$ and has a large ordered magnetic moment
$\mu=0.85\mu_{{\rm B}}$ on uranium atoms. 
UNi$_2$Al$_3$ orders 
in a spin-density wave (SDW) with an incommensurate wave vector ${\bf k}=
(H\pm0.61,0,0.5)$ and 
has a tiny magnetic moment of $0.2\mu_{{\rm B}}$ whose magnitude is 
modulated within the basal plane.
Regarding the superconducting properties,
various experiments suggest that UPd$_2$Al$_3$ is a spin-singlet
d-wave superconductor with a line-node gap. On the other hand, 
spin-triplet superconductivity in UNi$_2$Al$_3$ has been revealed by the NMR
measurement recently~\cite{rf:Ishida}.
We have proposed a theory of the superconductivity 
in UPd$_2$Al$_3$~\cite{rf:NiU}. Therefore, we have to explain 
the mechanism of the spin-triplet superconductivity in UNi$_2$Al$_3$ and 
the difference between the superconducting state
of UNi$_2$Al$_3$ and that of UPd$_2$Al$_3$ 
from the same viewpoint as that of our previous work,
although the detailed electronic structure of UNi$_2$Al$_3$ has not been
investigated yet.
Motivated by the point mentioned above, 
we discuss the possibility of spin-triplet superconductivity 
in a two-dimensional Hubbard model on a triangular 
lattice, as a first step.
In the case of spin-triplet superconductor ${\rm Sr_{2}RuO_{4}}$, 
Nomura and Yamada have recognized that 
the momentum and frequency dependence of 
the effective interaction between electrons, 
which is not included in the interaction mediated by the spin
fluctuation, is essential for realizing the spin-triplet pairing and 
they have explained the superconducting mechanism 
within the third-order perturbation theory (TOPT)~\cite{rf:No}.
The perturbation approach is sensitive to the
dispersion of the bare energy band, by its nature.
It implies that the lattice structures and the band
filling play essential roles in the calculation of superconducting
transition temperature $T_{{\rm c}}$.
Therefore, it is important to evaluate superconducting
transition temperature $T_{{\rm c}}$ 
on the basis of the detailed electronic structure in each system.
In this paper, we also calculate $T_{{\rm c}}$ of 
spin-triplet superconductivity in 
a two-dimensional Hubbard model on a triangular lattice within the TOPT.
In a similar model, Kuroki and Arita have proposed that
spin-fluctuation-mediated spin-triplet superconductivity can be realized 
in their model with disconnected Fermi surfaces, by using
fluctuation-exchange approximation (FLEX)~\cite{rf:KA}.
However, as mentioned later, our proposed 
mechanism of spin-triplet superconductivity is completely
different from their proposed mechanism.

We start from the quasi-particle state according to the discussion of 
our previous work~\cite{rf:NiCe}. We write the model Hamiltonian as follows:
\begin{eqnarray}
H&=&H_{0}+H_{1},\\
H_{0}&=&\sum_{{\bf k},\sigma}\left(\epsilon
({\bf k})-\mu
\right)
a_{{\bf k}\sigma}^{\dagger}a_{{\bf k}\sigma},\\
H_{1}&=&\frac{U}{2N}
\sum_{\sigma\neq\sigma^{\prime}}
\sum_{{\bf k}_{1}{\bf k}_{2}{\bf k}_{3}
{\bf k}_{4}}
\delta_{{\bf k}_{1}+{\bf k}_{2},
{\bf k}_{3}+{\bf k}_{4}}
a_{{\bf k}_{1}\sigma}^{\dagger}
a_{{\bf k}_{2}\sigma^{\prime}}^{\dagger}
a_{{\bf k}_{3}\sigma^{\prime}}a_{{\bf k}_{4}\sigma},
\end{eqnarray}
where $a_{{\bf k}\sigma}^{\dagger}(a_{{\bf k}\sigma})$
is the creation(annihilation) operator for the electron with
momentum ${\bf k}$ and spin index $\sigma$; 
$\epsilon({\bf k})$ and $\mu$ are the dispersion 
of the bare energy band on a two-dimensional 
triangular lattice and the chemical potential, 
respectively. 
The sum over ${\bf k}$ indicates taking summation over a
primitive cell of the inverse lattice.
In the above equations, we have rescaled the 
length, energy, temperature and time by
$a, t, \frac{t}{k_{\rm B}}, \frac{\hbar}{t}$.
(where $a, t, k_{\rm B}$ and $\hbar$ are the  lattice constant, 
the nearest neighbor hopping integrals, 
Boltzmann constant and Planck constant divided by $2\pi$, respectively)

\begin{figure}
\includegraphics[width=0.6 \linewidth]{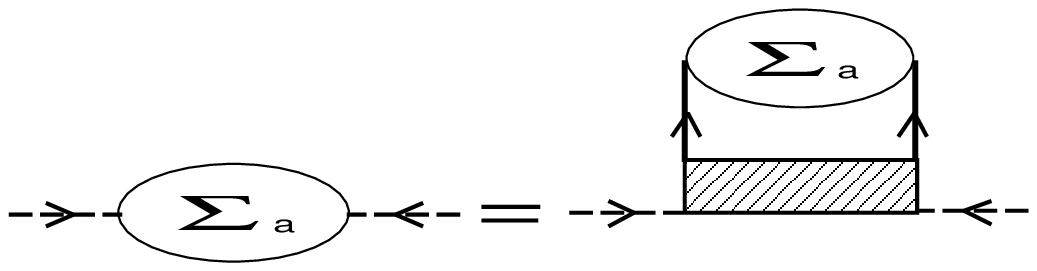}
\caption{\'Eliashberg's equation. 
The thick line represents Green's function
 with self-energy correction. The shaded rectangle represents
 the effective interaction.}
\label{fig:Eliash}
\end{figure}

\begin{figure}
\includegraphics[width=0.6 \linewidth]{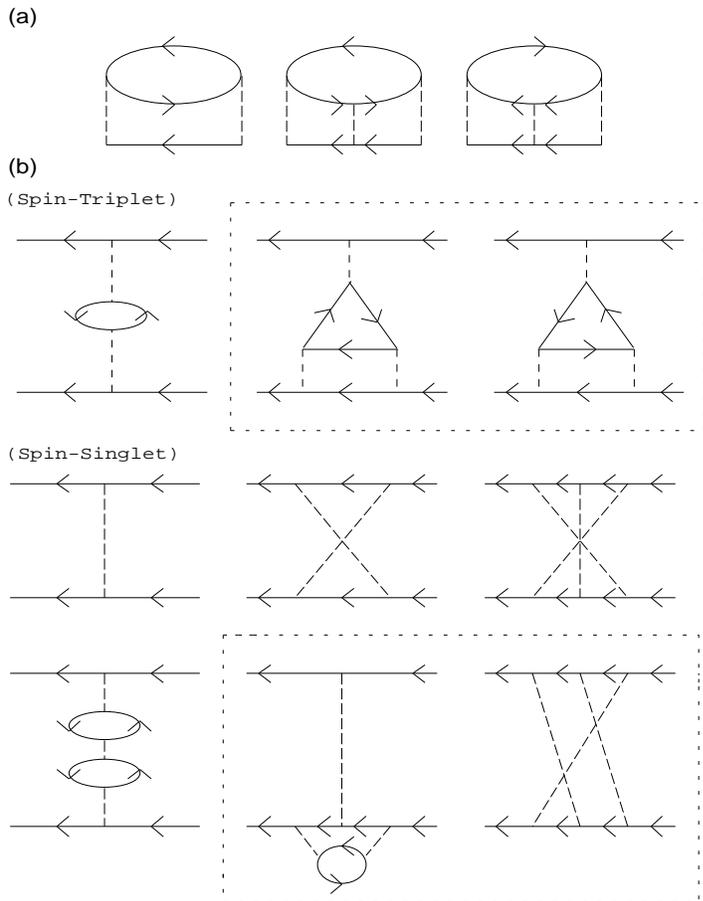}
\caption{(a) Feynman diagrams of the
 normal self-energy up to the third order. 
(b) Feynman diagrams of the effective interaction up to the third
 order. Solid and dashed lines correspond to the bare Green's function 
 and the interaction, respectively.}
\label{fig:Fey}
\end{figure}

We calculate $T_{{\rm c}}$ by solving 
 \'Eliashberg's equation (Fig.~\ref{fig:Eliash}).
In the equation, 
the normal self-energy and the effective interaction 
are obtained within the third-order perturbation 
with respect to $U$(Fig.~\ref{fig:Fey}).
The diagrams enclosed by a dashed line in Fig.~\ref{fig:Fey}(b) are 
the vertex correction terms which are not direct contributions from spin
fluctuations. The other diagrams are included in RPA.
We call the latter 'RPA-like diagrams' in this paper.
In Fig.~\ref{fig:Fey}(b), we omit writing 
the diagrams given by turning the 
vertex correction terms in Fig.~\ref{fig:Fey}(b) upside down.

Our model parameters are the dispersion $\epsilon({\bf k})$ of 
the bare energy band on a two-dimensional triangular lattice, the electron 
number $n$ per one spin site and the Coulomb repulsion $U$.
Regarding the dispersion of the bare energy band, we consider the following
dispersions of the bare energy band.

\begin{figure}
\includegraphics[width=0.6 \linewidth]{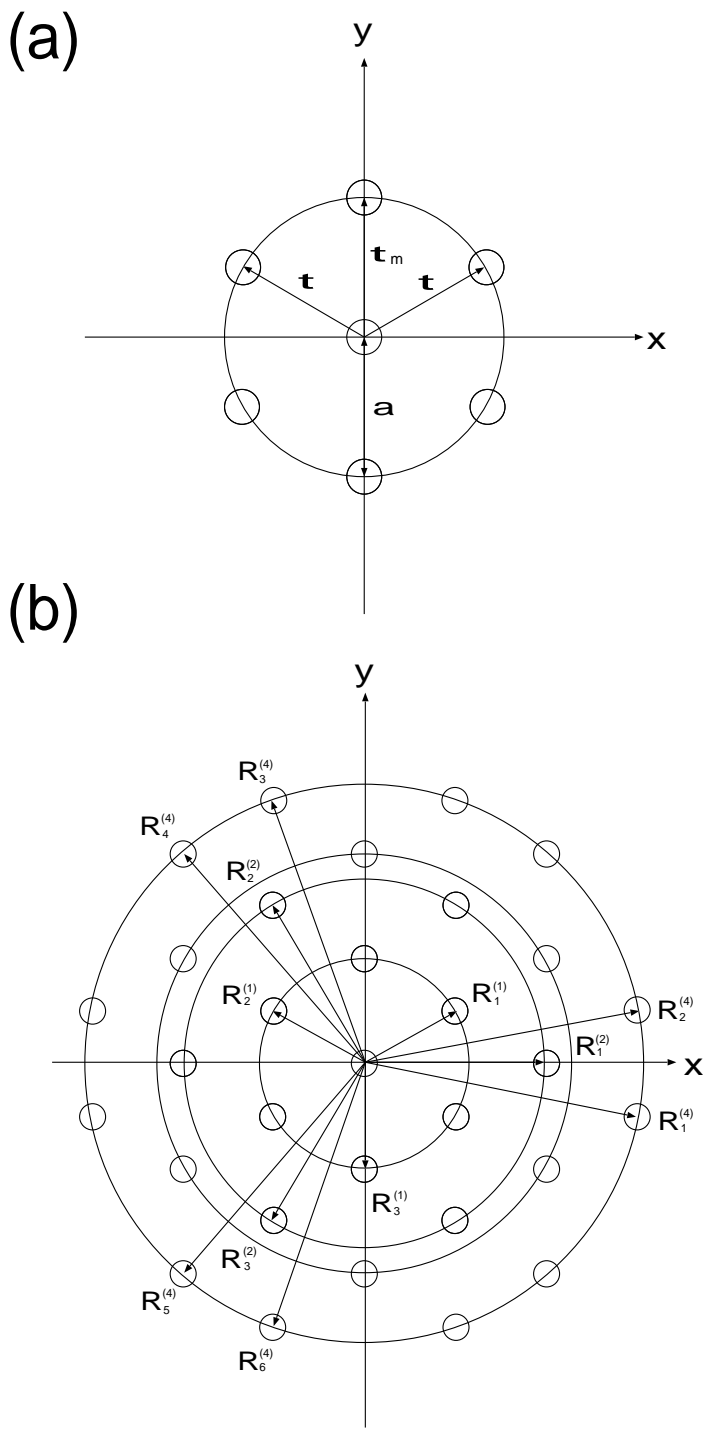}
\caption{(a) Hopping integrals with anisotropic nature (D$_2$-symmetry) on
 triangular lattice. (b) Labels of basis functions.}
\label{fig:D2latRep}
\end{figure}

At first, we consider the following:
\begin{equation}
\epsilon_{{\rm D}_{2}}({\bf k},t_{{\rm m}})=-4\cos(\frac{\sqrt{3}}{2}k_{x})\cos(\frac{1}{2}k_{y})-2t_{\rm m}\cos(k_{y})  ;t_{{\rm m}}\neq t(=1).
\end{equation}
$\epsilon_{{\rm D}_{2}}({\bf k},t_{{\rm m}})$ exhibits only the 
D$_2$-symmetry because of $t_{{\rm m}}\neq 1$ (Fig.~\ref{fig:D2latRep}(a)).
In the case of spin-singlet superconductor UPd$_2$Al$_3$, 
we adopted the above dispersion of the bare energy band in the
previous work~\cite{rf:NiU}. In the previous work, 
we have considered only the nearest neighbor
hopping integrals and we have assumed that the value of 
hopping integral $t_{\rm m}$ along the magnetic moment 
is different from the value of other hopping integrals $t$, because 
the superconductivity of ${\rm UPd_{2}Al_{3}}$ is realized in
the antiferromagnetic state.
Thus we have included the effect of the antiferromagnetic
order in the difference between $t_{\rm m}$ and $t$. We have also determined
the values so as to reproduce the considered Fermi sheet which is
obtained by the band calculation and is not of D$_6$-symmetry, 
reflecting the antiferromagnetic structure.
We have concluded that the main origin of the superconductivity is 
the momentum dependence of
the spin fluctuations which stems from the shape of our considered Fermi
sheet which undergoes symmetry breakdown(D$_6$ $\rightarrow$ D$_2$-symmetry) 
due to the antiferromagnetic order and then possesses nesting properties.
In the present paper, we investigate how superconducting states change
when we vary $t_{{\rm m}}$(symmetry in the system) 
from $t_{{\rm m}}\neq 1$(D$_2$) to $t_{{\rm m}}=1$(D$_6$).
We also investigate the possibility of spin-triplet 
superconductivity near the previous model of UPd$_2$Al$_3$.
Then we discuss a possible model of UNi$_2$Al$_3$, 
although the detailed electronic structure of UNi$_2$Al$_3$ has not been
investigated yet. 

Next we consider D$_6$-symmetric dispersions of the bare energy
band ({\it i.e}, dispersions of the bare energy band without 
anisotropic nature in the hopping integrals).
In this paper, we consider the nearest neighbor hopping model. We also 
consider the following dispersion of the bare energy band:
\begin{equation}
\epsilon_{{\rm D}_{6}}({\bf k})=\epsilon_{{\rm D}_{2}}({\bf k},1).
\end{equation}

To satisfy 
Luttinger's theorem, that is, the conservation law of particle number, 
we adjust the chemical potential $\mu$ by using the secant method.
To solve ${\rm\acute{E}}$liahberg's equation by using the
power method algorithm, we have to calculate the 
summation over the momentum and the frequency space. Since all summations
 are in the convolution forms, we can carry them out by using the
 algorithm of the fast Fourier transformation.
For the frequency, irrespective of the temperature, 
we have 1024 Matsubara
frequencies. Therefore, we calculate throughout 
in the temperature region $T\ge T_{\rm lim}$
, where $T_{\rm lim}$ is the lower limit temperature
 for reliable numerical calculation,
which is estimated to be about $3.0\times 10^{-3}
(>\Delta\epsilon/(2\pi\times 1024)\simeq 1.4\times 10^{-3})$, where
$\Delta\epsilon$ is the bandwidth; 
we divide a primitive cell into 128$\times$128 meshes.

To investigate how superconducting states change 
when we vary $t_{{\rm m}}$ (symmetry in the system) 
from $t_{{\rm m}}\neq 1$(D$_2$) to $t_{{\rm m}}=1$(D$_6$), 
we calculate the eigenvalues of ${\rm\acute{E}}$liahberg's equation for 
various $t_{{\rm m}}(0.75\sim 0.99)$ and $n$, 
starting from the model parameters of our previous work
($\epsilon_{{\rm D}_{2}}({\bf k},t_{{\rm m}}),t_{{\rm m}}=0.75, n=0.572$).
In this calculation, we select the Coulomb repulsion $U=7.5$.
In the case of the D$_2$-symmetric system, we can classify the eigenvalues
of ${\rm\acute{E}}$liahberg's equation according to the 
irreducible representations 
of D$_2$. D$_2$ has four irreducible representations (see Table~\ref{D2}).
\begin{table}
\caption{Irreducible representations of D$_2$}
\label{D2}
\begin{tabular}{@{\hspace{\tabcolsep}\extracolsep{\fill}}ccc} \hline
Irreducible representation & Parity  &Basis functions (maximum wavelength in ${\bf k}$-space)\\ \hline
A        & even &1 \\
B$_1$    & even & $c^{(1)}_{1}- c^{(1)}_{2}$\\ \hline
B$_2$    & odd  & $s^{(1)}_{1}+s^{(1)}_{2},s^{(1)}_{3}$ \\
B$_3$    & odd  & $s^{(1)}_{1}- s^{(1)}_{2}$\\ \hline
\end{tabular}
\end{table}

The short notations $s^{(i)}_{j}$ and $c^{(i)}_{j}$ in 
Tables~\ref{D2} and ~\ref{D6} mean that $s^{(i)}_{j}\equiv\sin({\bf
R}^{(i)}_{j}\cdot{\bf k})$ and 
$c^{(i)}_{j}\equiv\cos({\bf R}^{(i)}_{j}\cdot{\bf k})$
, where ${\bf R}^{(i)}_{j}$ is defined in Fig.~\ref{fig:D2latRep}(b).

The eigenvalues which belong to A or B$_1$ correspond to spin-singlet
states. The eigenvalues which belong to B$_2$ or B$_3$ correspond 
to spin-triplet states. 
Based on the values of our parameters, 
in the case of spin-singlet states, the maximum eigenvalue belongs to
B$_1$ and in the case of spin-triplet states, the maximum eigenvalue
belongs to B$_2$.

\begin{figure}
\includegraphics[width=1 \linewidth]{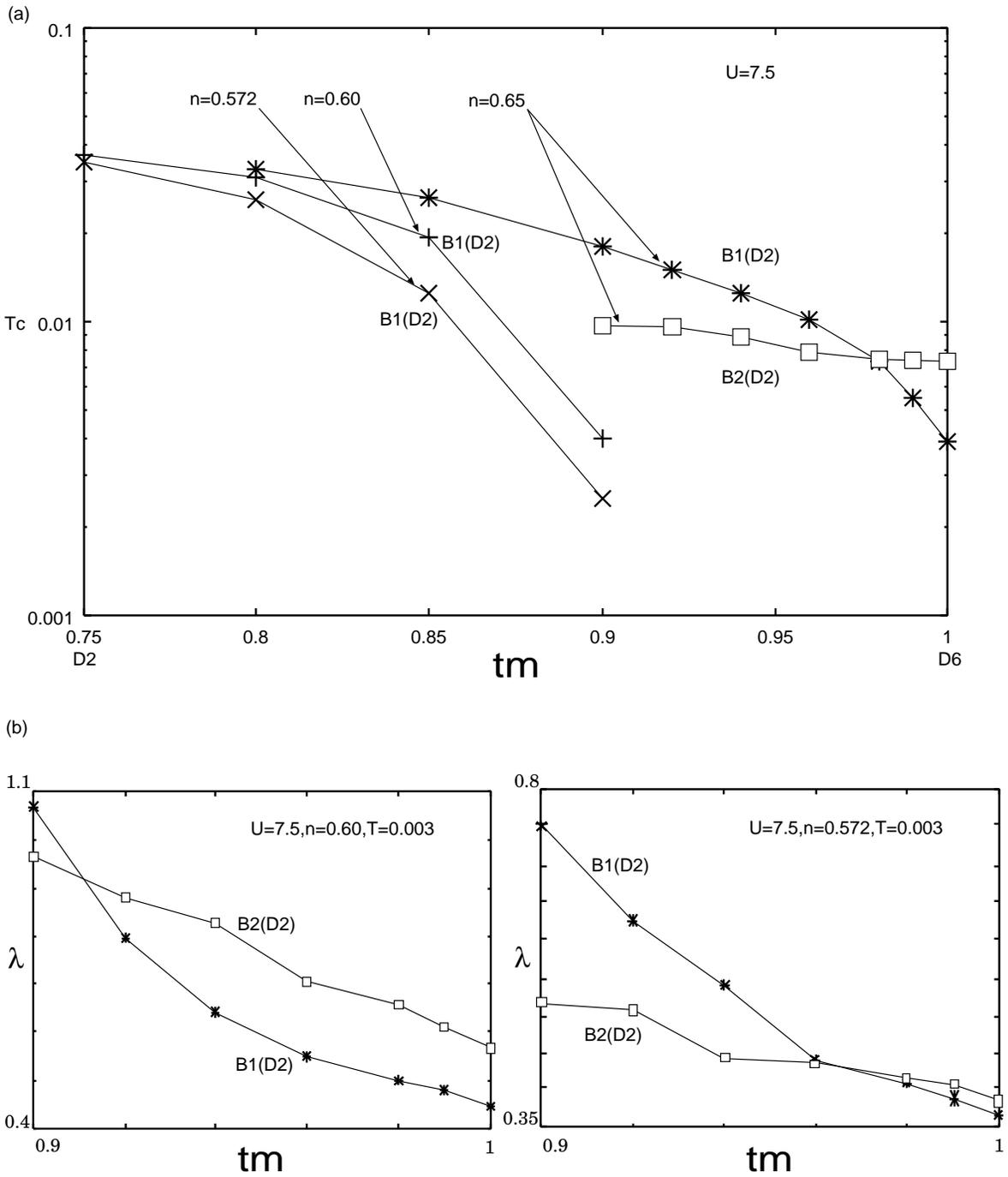}
\caption{(a) The calculated 
$T_{{\rm c}}$ is shown as a function of $t_{{\rm m}}$ 
for various electron numbers $n$ per one spin site. (b) The
 calculated eigenvalue is shown as a function of $t_{{\rm m}}$ at
 $T_{{\rm lim}}=0.003$ for $n=0.572$ and $0.600$}
\label{fig:TcD2}
\end{figure}

The results are shown in Fig.~\ref{fig:TcD2}.
From this figure, the following facts are evident.

In the case of $n=0.572$ and $t_{{\rm m}}=0.75$,
 the spin-singlet state is most stable. 
We have pointed out that this spin-singlet state corresponds 
to that of UPd$_2$Al$_3$ in our
previous work.
When we vary $t_{{\rm m}}$ (symmetry in the system) 
from $t_{{\rm m}}=0.75$(D$_2$) to $t_{{\rm m}}=1$(D$_6$), 
at the same time, the anisotropic nature in spin fluctuation is 
suppressed. In this case, the spin-singlet state 
is suppressed because the main origin of the d-wave superconductivity
is the momentum and frequency dependence of  
spin fluctuations as we have pointed out in our previous work, 
and we can see that spin-triplet states have the tendency to emerge.

Next, we consider D$_6$-symmetric dispersions $\epsilon_{{\rm
D}_{6}}({\bf k})$ of the bare energy band and calculate $T_{{\rm c}}$ 
for some model parameters $U$ and $n$.
In the case of the D$_6$-symmetric system, we can classify the eigenvalues
of ${\rm\acute{E}}$liahberg's equation according to the 
irreducible representations 
of D$_6$. D$_6$ has six irreducible representations (see Table~\ref{D6}).
\begin{table}
\caption{Irreducible representations of D$_6$}
\label{D6}
\begin{tabular}{@{\hspace{\tabcolsep}\extracolsep{\fill}}ccc} \hline
Irreducible representation & Parity  &Basis functions (maximum
 wavelength in ${\bf k}$-space)  \\ \hline
A$_1$    & even  & 1 \\
A$_2$    & even  & $c^{(4)}_{1}-c^{(4)}_{2}+c^{(4)}_{3}-c^{(4)}_{4}+c^{(4)}_{5}-c^{(4)}_{6}$\\
E$_2$    & even  & $\{c^{(1)}_{1}-c^{(1)}_{2},c^{(1)}_{1}+c^{(1)}_{2}-2c^{(1)}_{3}\}$\\ \hline
B$_1$    & odd   & $s^{(1)}_{1}+s^{(1)}_{2}+s^{(1)}_{3}$\\
B$_2$    & odd   & $s^{(2)}_{1}+s^{(2)}_{2}+s^{(2)}_{3}$\\
E$_1$    & odd   & $\{s^{(1)}_{1}-s^{(1)}_{2},s^{(1)}_{1}+s^{(1)}_{2}-2s^{(1)}_{3}\}$\\ \hline
\end{tabular}
\end{table}

The eigenvalues which belong to A$_1$, A$_2$ or E$_2$ correspond to spin-singlet states. 
The eigenvalues which belong to B$_1$, B$_2$ or E$_1$ correspond 
to spin-triplet states. 
Based on the values of our parameters, 
the maximum eigenvalue belongs to E$_1$ 
which corresponds to a spin-triplet state.  

The dependences of $T_{{\rm c}}$ on $U$ and $n$ are shown
Fig.~\ref{fig:TcD6}.
On the basis of these results, we can point out the following
findings.
For large $U$, high $T_{\rm c}$ values are
obtained for all parameters.

When we increase the electron density $n$ for the fixed valued
 Coulomb repulsion $U$ and the Fermi level becomes close to the
van Hove singularity (whose results are not presented in this paper), 
then  high $T_{\rm c}$ values are obtained.

To examine how the vertex corrections influence  $T_{\rm c}$,
we also tried to calculate $T_{\rm c}$ by including only the 
RPA-like diagrams but we 
could not find any finite value of $T_{{\rm c}}$ within the
precision of our numerical calculations.
Therefore, we can calculate only the eigenvalue 
by including only the RPA-like diagrams at $T_{{\rm c}}$ obtained 
by using the TOPT.

From these results, it is found that the vertex terms, 
which are not included in the interaction mediated by the spin
fluctuation, are essential for realizing the spin-triplet pairing.

In this paper, we have discussed
 the possibility of spin-triplet superconductivity 
in a two-dimensional Hubbard model on a triangular 
lattice within the TOPT and we have obtained
spin-triplet superconducting states in near a D$_6$-symmetric system.
In this case, the vertex correction terms play an essential role
in obtaining the spin-triplet superconducting states.
This is the same result that Nomura and Yamada have obtained 
in the investigation of the mechanism of spin-triplet superconductor
 Sr$_2$RuO$_4$ within the TOPT, as we have mentioned above.
Jujo {\it et al.} have investigated the mechanism of the spin-singlet 
superconductivities in organic superconductors $\kappa$-type
(BEDT-TTF)$_2$X within the TOPT and FLEX~\cite{rf:Ju}.
They have compared the results between the TOPT and FLEX, and have 
pointed out that the vertex correction terms (which are also not included
in the FLEX) have a crucial effect on the calculation of 
$T_{{\rm c}}$ for strongly frustrated systems. 
Therefore, our presented results are consistent with the above two results.
Recently, 
Nisikawa {\it et al.} have discussed the spin-singlet superconductivity 
in CeIr$_x$Co$_{1-x}$In$_5$ within the 
TOPT and have pointed out that the vertex 
correction terms are important to explain 
the $x$-dependence of $T_{{\rm c}}$~\cite{rf:NiCe}.
The findings described 
above suggest that the calculations of $T_{{\rm c}}$, 
which include only the spin fluctuation terms, are questionable and
 should be carefully performed with vertex corrections.

UNi$_2$Al$_3$ and UPd$_2$Al$_3$ undergo superconducting transition 
below the N${\rm\acute{e}}$el temperature.
Therefore, the symmetry in the itinerant electron system under the 
antiferromagnetic structure is one of the important matters 
to consider in investigating the mechanism of superconductivity. 
In the case of UPd$_2$Al$_3$, the symmetry in the itinerant electron
system is not of hexagonal symmetry, reflecting the effect of 
the antiferromagnetic structure with a large ordered magnetic moment
$\mu=0.85\mu_{{\rm B}}$ on uranium atoms, and we have treated this 
by considering the anisotropic hopping integral $t_{{\rm m}}\neq 1$.
In this context, the symmetry in the itinerant 
electron system of UNi$_2$Al$_3$ 
may be  more isotropic than the symmetry in that of UPd$_2$Al$_3$ because
of a reflecting incommensurate 
SDW order with a tiny moment of 0.2$\mu_{{\rm B}}$, 
although the detailed electronic structure of UNi$_2$Al$_3$ has not been 
investigated yet. 
Therefore, we assume that
$t_{{\rm m}}\simeq1$ in the case of UNi$_2$Al$_3$. 
Based on the hypothesis mentioned above, 
our results seem to explain not only the
mechanism of spin-triplet superconductivity in UNi$_2$Al$_3$ but also the
difference between the superconductivity of UNi$_2$Al$_3$ and that of 
UPd$_2$Al$_3$ because the spin-singlet superconductivity in 
the D$_2$-symmetric system($t_{{\rm m}}\neq 1$) is suppressed toward 
D$_6$-symmetry and the spin-triplet
superconductivity emerges in near the D$_6$-symmetric system ($t_{{\rm m}}=1$).

In conclusion, we discussed
 the possibility of spin-triplet superconductivity 
in a two-dimensional Hubbard model on a triangular 
lattice within the TOPT. 
We obtained spin-triplet superconducting states in near 
the D$_6$-symmetric system.
We pointed out  the possibility 
that our obtained results correspond to the
difference between the superconductivity of UNi$_2$Al$_3$ and that of 
UPd$_2$Al$_3$.

Numerical computation in this work was carried out at the Yukawa
 Institute Computer Facility.

\newpage

\begin{figure}
\includegraphics[width=1 \linewidth]{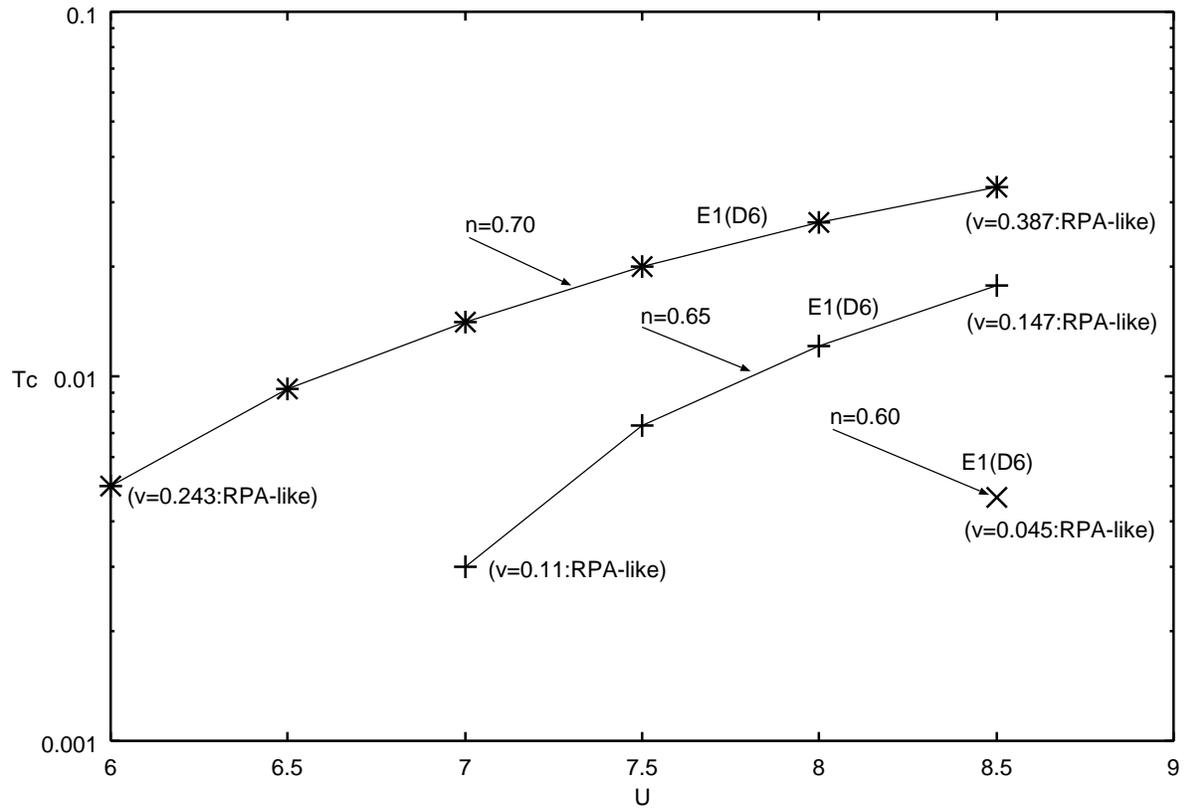}
\caption{The calculated $T_{{\rm c}}$ is shown as a function of $U$ 
for various electron numbers $n$ per one spin site.''
In this figure, 
(v=***:RPA-like)'' means that eigenvalue v calculated by including only
 the RPA-like diagrams is v=*** at $T_{{\rm c}}$ obtained by using the TOPT.}
\label{fig:TcD6.eps}
\end{figure}

\end{document}